\documentclass[pmlr,twocolumn,10pt]{jmlr} 

\usepackage{longtable}

\usepackage{booktabs}
\usepackage{graphicx}
\usepackage[normalem]{ulem}
\useunder{\uline}{\ul}{}

\usepackage[switch]{lineno}

\usepackage{xcolor}

\theorembodyfont{\upshape}
\theoremheaderfont{\scshape}
\theorempostheader{:}
\theoremsep{\newline}

\jmlryear{2024}
\jmlrworkshop{Conference on Health, Inference, and Learning (CHIL) 2024} 

\title[SQUWA: Signal Quality Aware DNN Architecture]{SQUWA: Signal Quality Aware DNN Architecture for Enhanced Accuracy in Atrial Fibrillation Detection from Noisy PPG Signals}

\author{%
 \Name{Runze Yan} \Email{runze.yan@emory.edu}\\
 \addr Center for Data Science, Nell Hodgson Woodruff School of Nursing, Emory University\\
 \Name{Ding Cheng} \Email{chengding@gatech.edu}\\
 \addr Department of Biomedical Engineering, Georgia Institution of Technology \& Emory University\\
 \Name{Ran Xiao} \Email{ran.xiao@emory.edu}\\
 \addr Center for Data Science, Nell Hodgson Woodruff School of Nursing, Emory University\\
 \Name{Aleksandr Fedorov} \Email{aleksandr.vladimirovich.fedorov@emory.edu}\\
 \addr Center for Data Science, Nell Hodgson Woodruff School of Nursing, Emory University\\
 \Name{Randall J Lee} \Email{randall.lee@ucsf.edu}\\
 \addr School of Medicine, University of California at San Francisco\\
 \Name{Fadi B Nahab} \Email{fnahab@emory.edu}\\
 \addr Department of Neurology, School of Medicine, Emory University\\
 \Name{Xiao Hu} \Email{xiao.hu@emory.edu}\\
 \addr Center for Data Science, Nell Hodgson Woodruff School of Nursing, Emory University\\
}

\begin{document}

\maketitle

\begin{abstract}
Atrial fibrillation (AF), a common cardiac arrhythmia, significantly increases the risk of stroke, heart disease, and mortality. Photoplethysmography (PPG) offers a promising solution for continuous AF monitoring, due to its cost efficiency and integration into wearable devices. Nonetheless, PPG signals are susceptible to corruption from motion artifacts and other factors often encountered in ambulatory settings. Conventional approaches typically discard corrupted segments or attempt to reconstruct original signals, allowing for the use of standard machine learning techniques. However, this reduces dataset size and introduces biases, compromising prediction accuracy and the effectiveness of continuous monitoring. We propose a novel deep learning model, \textbf{\underline{S}}ignal \textbf{\underline{Qu}}ality \textbf{\underline{W}}eighted Fusion of \textbf{\underline{A}}ttentional Convolution and Recurrent Neural Network (SQUWA), designed to learn how to retain accurate predictions from partially corrupted PPG. Specifically, SQUWA innovatively integrates an attention mechanism that directly considers signal quality during the learning process, dynamically adjusting the weights of time series segments based on their quality. This approach enhances the influence of higher-quality segments while reducing that of lower-quality ones, effectively utilizing partially corrupted segments. This approach represents a departure from the conventional methods that exclude such segments, enabling the utilization of a broader range of data, which has great implications for less disruption when monitoring of AF risks and more accurate estimation of AF burdens. Moreover, SQUWA utilizes variable-sized convolutional kernels to capture complex PPG signal patterns across different resolutions for enhanced learning. Our extensive experiments show that SQUWA outperform existing PPG-based models, achieving the highest AUCPR of 0.89 with label noise mitigation. This also exceeds the 0.86 AUCPR of models trained with using both electrocardiogram (ECG) and PPG data.
\end{abstract}

\paragraph*{Data and Code Availability}
Research data will not be shared for ethical reasons, except for one publicly accessible. The detailed data description are in Section~\ref{sec:training_data}. Code is available at \url{https://github.com/Runz96/SQUWA}.

\section{Introduction}
\label{sec:intro}
Atrial Fibrillation (AF), the most common chronic cardiac arrhythmia, impacts around 33.5 million people globally, with its occurrence increasing~\cite{chugh2014worldwide}. Notably, AF significantly contributes to health risks, accounting for 20\% of all strokes and a third of all hospital admissions due to heart rhythm issues~\cite{marini2005contribution}. To reduce AF-associated risks, it is important to be able to detect AF early in the trajectory, allowing for timely intervention that can mitigate the progression of electrical and structural remodeling of atrial tissue~\cite{hart2007meta}. Although electrocardiogram (ECG) signals are the gold standard for detecting AF, their practical application is limited by challenges in long-term daily wearability, necessitating the exploration of alternative signal modalities to monitor and detect AF such as photoplethysmography (PPG) signals~\cite{charlton20232023}. PPG signals represent blood volume changes in the microvascular bed of tissue, providing a non-invasive method to capture the characteristics of irregular heart rhythms of AF. Their potential for AF detection, combined with their presence in about 71\% of consumer wearables has highlighted their significance~\cite{henriksen2018using}. However, the utility of PPG in AF detection is often undermined by noise such as motion artifacts~\cite{seok2021motion}. Thus, accurate classification of these noise-affected PPG signals is critical for the development of a system that is both highly sensitive and precise in detecting AF, in particular for screening AF at scale.

Despite advancements in AF detection through PPG data analysis using Deep Neural Networks (DNNs) such as Convolutional Neural Networks (CNN)~\cite{shashikumar2017deep}, Long Short-Term Memory (LSTM)~\cite{cheng2020atrial}, and Transformer Neural Networks~\cite{nankani2022atrial}, the issue of noise and motion artifacts in raw PPG signals remains unresolved. Current methods for the corrupted PPG signals involves discarding low-quality samples, or enhancing the signal-to-noise ratio (SNR), so that the standard neural network methods can focus on the 'clean' data and tend to ignore signals that are not of high quality~\cite{liaqat2020detection}. However, discarding low-quality signals can reduce the volume of data available for model training, leading to challenges in estimating metrics like AF burden~\cite{pereira2019supervised, zhu2021atrial}. This strategy often relies on arbitrary thresholds to remove low-quality signals~\cite{roy2020photoplethysmogram}. Other approaches attempt to improve signal quality before detecting AF, which can lead to errors accumulation and propagation from the enhancement stage to detection, potentially compromising the algorithm's effectiveness~\cite{ding2023log, afandizadeh2023accurate}. Conversely, a streamlined, one-step method that directly integrates PPG signal quality into the AF detection process could provide a more efficient solution. This method enhances accuracy by making the most of the data available during training, without altering the raw signal.


In this study, we introduce a novel DNN architecture, termed \textbf{\underline{S}}ignal \textbf{\underline{Qu}}ality \textbf{\underline{W}}eighted Fusion of \textbf{\underline{A}}ttentional Convolution and Recurrent Neural Network (SQUWA), designed for AF detection using PPG data.  Unlike conventional methods that exclude low-quality signals, SQUWA integrates an innovative attention mechanism that dynamically assigns weights to PPG segments based on their signal qualities. This mechanism is not an isolated feature but is embedded within the AF detection process itself. It's trained to incorporate signal quality levels directly into the learning model, moving away from the detached, two-step methods. In practice, during AF detection, SQUWA prioritizes segments with higher quality to have a greater influence on the prediction. Conversely, segments that are significantly affected by noise receive less weight. This approach allows SQUWA to maximize the use of high-quality segments in a PPG sample, thereby minimizing the impact of lower-quality segments on the overall analysis, ensuring the model's predictions are informed by the most reliable data available. Moreover, this attention mechanism operates at a detailed temporal resolution, processing each data point independently rather than treating an entire PPG sample as a homogeneous unit. This refined approach enhances the overall effectiveness and accuracy of the AF detection process.

Additionally, to support this attention mechanism, we utilize  a class activation map (CAM)~\cite{zhou2016learning} derived from a pre-trained signal quality (SQ) model. This CAM generates a continuous signal quality index (SQI) for each PPG signal, offering a granular view of signal quality over time~\cite{pereira2019deep}.  Another novel feature of SQUWA is the initial decomposition of a PPG signal. This involves decomposing the raw signal and its first and second derivatives using a variety of CNN kernels, each with different kernel sizes. These decomposed elements are then strategically reassembled by a sub-network to create a composite signal. This approach is crucial in highlighting relevant signal features, significantly improving the ability to distinguish between AF and non-AF conditions. 


Our training and testing approach is meticulously designed to enhance the model's robustness and generalizability. We train the model on over 5 million PPG samples, ensuring comprehensive learning that encompasses a wide range of etiological variations within this disease population. For testing, we rigorously evaluate it on three external datasets. This strategy not only tests the model's adaptability to varied conditions but also confirms its effectiveness across diverse real-world scenarios. Experimental results show that our proposed approach enhances the accuracy of AF detection in the presence of varying signal quality. The SQUWA method outperforms baseline models, including CNN and RNN-based single-modality AF detection neural networks, across three external test sets. Remarkably, our method also shows competitive results when compared to an AF detection model trained using both PPG and ECG data. To our best knowledge, the proposed approach is the first deep learning framework that considers signal quality as an integral element in learning an AF detector. The contribution of this work can be summarized as follows: 
\begin{itemize}
    \item We introduced an attention mechanism that simultaneously considers the quality of PPG signals and AF detection. This method learns to weigh more on segments with higher SQIs, ensuring accurate AF classification even from partially compromised PPG signals. 
    \item The SQUWA model utilzed an adaptive attention sub-network to combines the raw PPG signal and its first and second derivatives. This approach allows the model to analyze the signal and its instantaneous rate of change and curvature, providing a comprehensive understanding of the signal dynamics.
    \item The proposed method outperforms the baseline PPG models and we were able to confirm through model interpretation analysis that the good quality segments in a PPG are indeed mapped to higher attention weights as learned by SQUWA.
\end{itemize}

\section{Related Work}
Recent progress in model design has focused on addressing the challenges posed by noisy signals. This includes strategies for both preprocessing—to improve signals before classification—and post-processing, aimed at refining predictions to offset the impact of partially corrupted data segments.~\cite{chatterjee2020review, zhang2021eeg}. The following paragraphs will highlight the strengths and weaknesses of common techniques within these two categories.

A notable preprocessing strategy for signal denoising involves the use of Variational Autoencoders (VAEs) and Generative Adversarial Networks (GANs) to augment data~\cite{im2017denoising, brophy2023generative}. These models have shown promise in restoring corrupted data of various types, such as images~\cite{im2017denoising}, acoustic signals~\cite{kuo2020dnae}, and text~\cite{zhu2018generative}. Denoising GAN (DN-GAN)~\cite{chen2020dn} and Denoising Autoencoders~\cite{bengio2013generalized} have been explored for their capacity to learn complex, high-level representations of data and for their ability to filter out noise, respectively. However, their success relies on the assumption that the noise patterns are predictable or follow certain distributions. This assumption may not hold in real-world scenarios where noise and corruption can be unpredictable and non-uniform. 

Transfer Learning has also been utilized to mitigate the challenges of partially corrupted signals~\cite{pan2009survey}. \cite{zhang2020learning} leveraged a pre-trained models on a clean speech dataset and adapted them to specific speech recognition task with lower quality data. ~\cite{kim2020transfer} has  utilized transfer learning to address the challenges by training a network with synthetic noise and then transferring this knowledge to effectively handle the varying characteristics of real-world noise. However, the effectiveness of transfer learning is limited by the availability of representative training data and the discrepancies between training and application datasets~\cite{day2017survey}. Models trained on high-quality data often struggle to adapt to lower quality signals, leading to challenges in accurately interpreting biomedical signals like PPG-based AF detection~\cite{pan2009survey}. Significant discrepancies in data quality and patient demographics often require time-consuming fine-tuning and domain adaptation.


Attention mechanisms are designed to direct a model's focus to the most significant parts of the input, improving learning efficiency and accuracy. ~\cite{niu2021review}. Attention methods have been successfully applied in many tasks, e.g., machine translation~\cite{tan2020data}, computer vision~\cite{guo2022attention}, and even physiological signals for AF detection~\cite{mousavi2020han}. A closely related study introduced a unique heart rate estimation method that combines a signal quality attention mechanism with an LSTM network~\cite{gao2022remote}, which proposed a novel remote heart estimation algorithm from video. However, this approach only uses attention mechanisms for offline correction of heart rate data, not for integrating signal quality into the learning phase. While attention mechanisms can aid in leveraging high-quality segments and reducing biases from low-quality ones, their use in addressing the challenges of making predictions from partially corrupted data remains unexplored.

\section{Methods}
Figure~\ref{fig:gen_frame} visualizes the overall structure of SQUWA, which begins by generating a composite signal from the raw PPG and its first and second derivatives. This process involves the use of variable-sized kernels to break down these signals. Subsequently, an attention subnet aggregates the kernel outputs through a weighted sum. A deep CNN then processes this composite signal to extract features with a lower temporal dimension to facilitate the subsequent temporal integration through a LSTM. Moreover, a pre-trained CNN-based signal quality (SQ) model evaluates the raw PPG signal, producing a signal quality index (SQI) over time. These LSTM outputs and SQIs are then combined through a signal quality attention (SQ-attention) mechanism. This nuanced integration of both signal features and SQIs weighs more contributions from locations in the signal where SQIs are high so that valuable information from partially corrupted PPG signals is effectively utilized for accurate classification. End-to-end training of SQUWA enables this data-driven integration process to maximize sensitivity and minimize false detection.

\begin{figure}[!ht]
\floatconts
  {fig:gen_frame}
  {\caption{Structure of Signal Quality Weighted Fusion of Attentional Convolution and Recurrent Neural Network (SQUWA).}}
  {\includegraphics[width=0.85\linewidth]{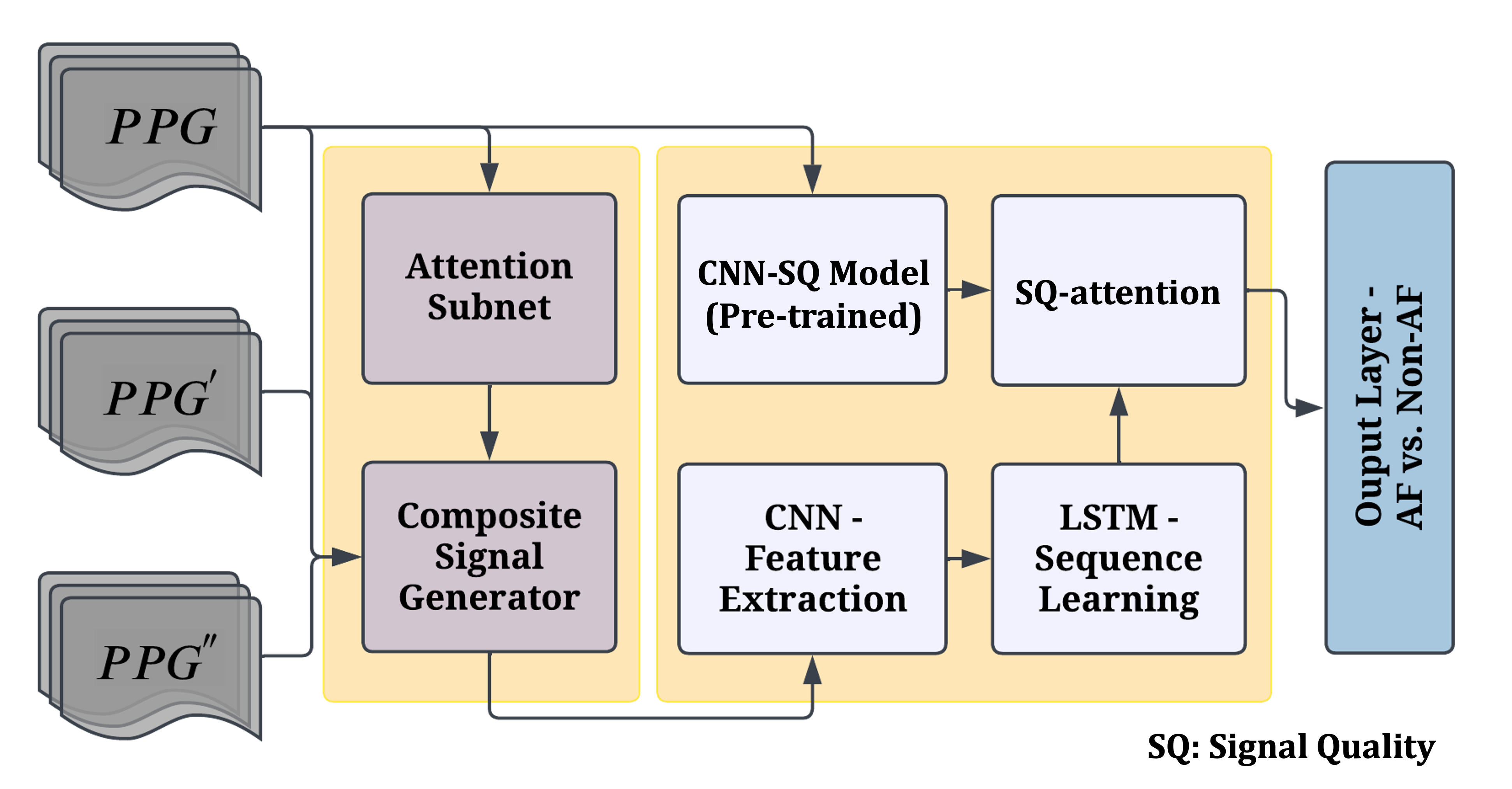}}
\end{figure}

\subsection{Composite Signal Generation}
The composite signal generation incorporates the two components found within the yellow box on the left side of Figure~\ref{fig:gen_frame}. The inputs of the SQUWA network are the raw PPG signal and its first and second derivatives. A PPG signal captures the heart's pulse by tracking changes in blood volume with each beat. The first derivative of the PPG signal relates to the velocity of the blood flow, indicating the rapidity of blood volume changes within the vessels and mirroring the pulse's rhythm. The second derivative offers a deeper look at how quickly the blood flow's velocity changes, essentially gauging the acceleration or deceleration of blood within the vessels. As shown in Figure~\ref{fig:signal_compositor}, these three forms of the PPG signal are analyzed in parallel by convolutional kernels of three different lengths ranging from a short to a long scale. The exact length of the kernel and the number of kernels will be fine-tuned during the training process, but the goal is to decompose the input at different scales and learn how to selectively combine them into a composite signal. The output from all kernels of the same length will be combined by using a 1x1  kernel, and this process results in nine component signals of the same length as that of the original PPG. An attention subnet learns weights to combine these nine components in a way that is determined by the characteristics of raw PPG. This attention subnet includes a convolutional layer, a fully-connected layer, and a SoftMax layer.

\begin{figure}[!ht]
\floatconts
  {fig:signal_compositor}
  {Visualization of the composite signal generation in the proposed SQUWA algorithm.}
  {\includegraphics[width=0.95\linewidth]{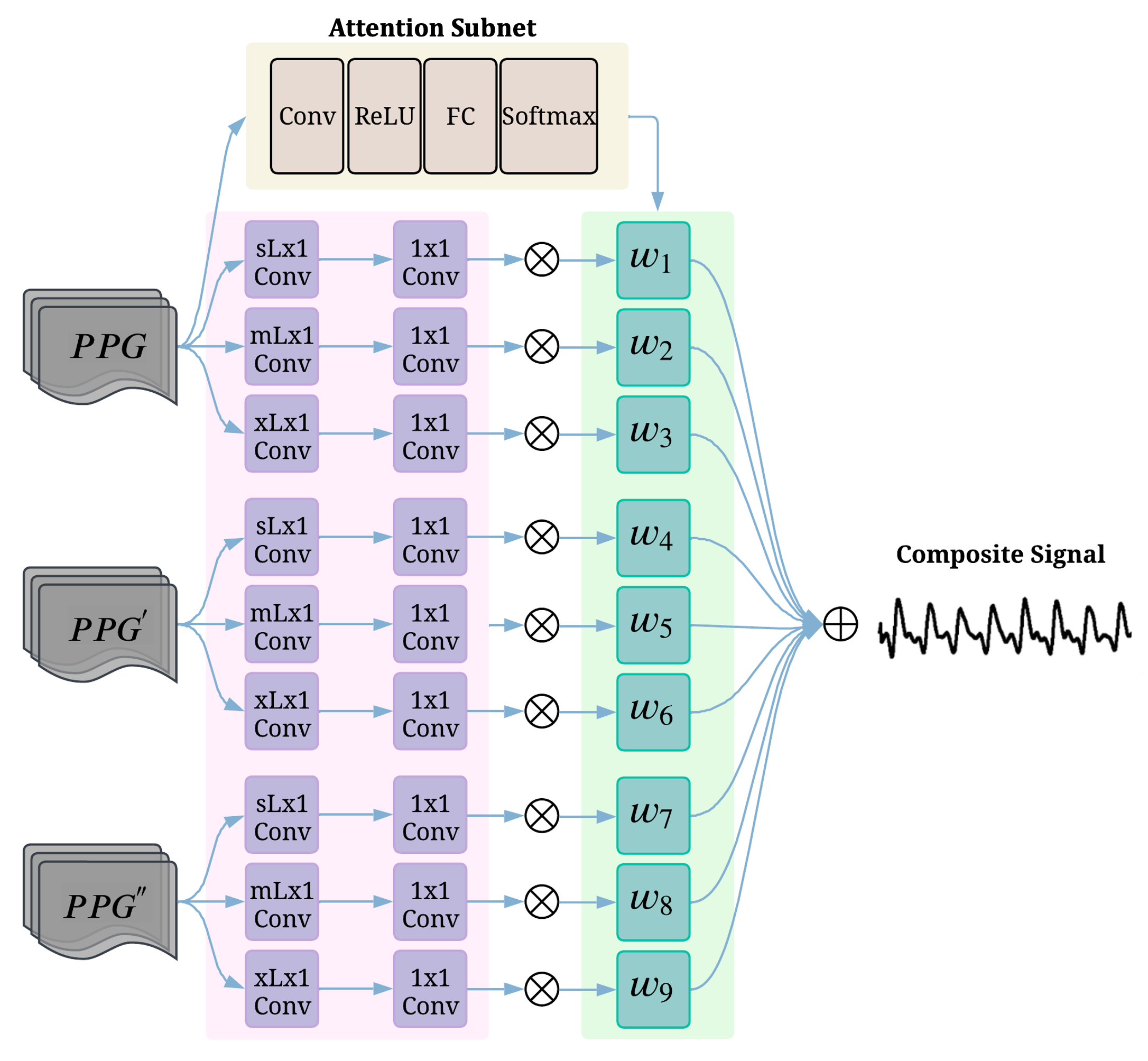}}
\end{figure}

\subsection{CNN-LSTM Fusion}
Figure~\ref{fig:rnn_cnn} highlights two branches: one for processing the composite signal and the other for processing the raw PPG. The composite signal is analyzed by a CNN to effectively extract a sequence of feature vectors. These vectors have a dimension of $n \times T$,  where $n$ is the number of kernels in the last convolutional layer and $T$ is the temporal dimension of the sequence. Because of pooling layers in the CNN, $T$ will be smaller than the length of the original signal. The second branch uses a CNN-based SQ model that processes raw PPG signals to produce SQIs. This SQ model, pre-trained on a small PPG dataset labeled with good and bad signal quality, differentiates between these qualities. We utilized the class activation map (CAM) from the last layer (prior to a global average pooling layer) of the SQ model as SQIs to reflect PPG quality over time~\cite{zhang2021explainability}. To match the temporal dimension with the feature-extraction CNN, we make sure the down-sampling factors in both CNN networks are identical so that each element of a feature sequence from the first branch can be characterized by a scalar SQI. We use 1-D ResNet as the backbone for the CNN feature extraction layer and SQ model and leave the exploration of other more modern CNN architecture for future work~\cite{alzubaidi2021review}. These two networks do not need to have identical architectures but need to have the same downsampling  factors to be temporally in sync. The output of the CNN feature extractor will be processed by a LSTM, and we use the one-directional LSTM as the backbone. The output from the LSTM has the dimension of $k \times T$, where $k$ is the number of hidden units in the LSTM layer.

\begin{figure}[!ht]
\floatconts
  {fig:rnn_cnn}
  {\caption{Visualization of the CNN-LSTM fusion and SQ-attention process.}}
  {\includegraphics[width=0.95\linewidth]{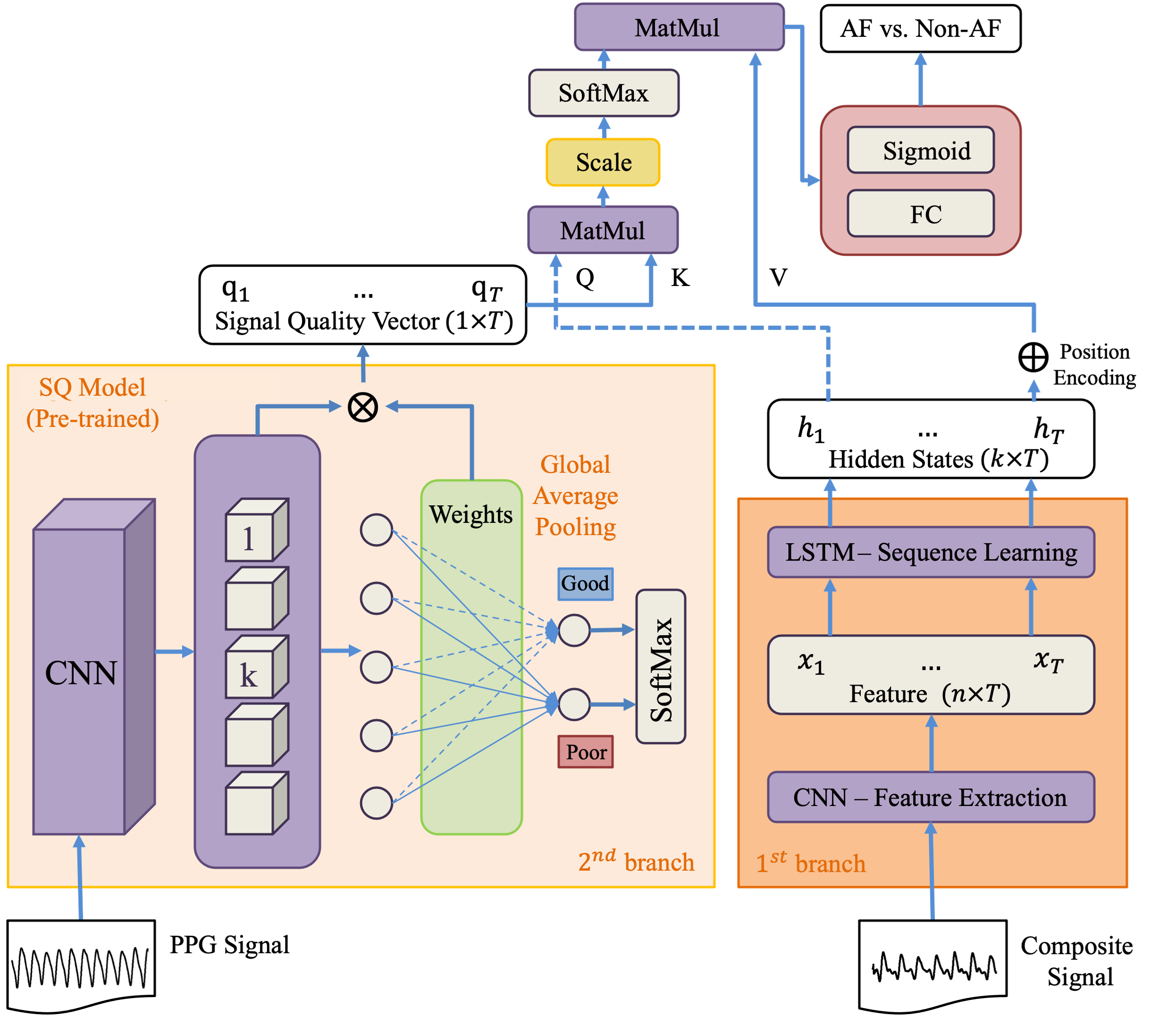}}
\end{figure}

\subsection{Signal Quality (SQ) - Attention}
In this section, we introduced the SQ-attention mechanism that begins by analyzing the hidden states $H$ from the LSTM, which captures the temporal sequential patterns in the PPG signal, along with signal quality values $SQI$ that assess the quality of each segment of the PPG signal in Figure~\ref{fig:rnn_cnn}. As shown in Formula~\ref{eq:qkv}, the mechanism converts these hidden states $H$ and $SQI$ values into queries ($Q$), keys ($K$), and values ($V$), which are simplified representations to help the model determine the segments of the signal that should be prioritized. 
\begin{align}
H_{\text{adj}} &= H + P\nonumber\\
Q &= H_{\text{adj}}^T \cdot W_Q\nonumber\\
K &= SQI^T \cdot W_K\nonumber\\
V &= H_{\text{adj}}^T \cdot W_V
\label{eq:qkv}
\end{align}
, where $W_{Q}$, $W_{K}$, and $W_{V}$ are the weight matrices for the query, key, and value transformation, respectively. And $P$ represents the positional encoding.

The attention mechanism operates by pairing queries with keys. This is done by measuring how much each query matches with a key, taking the product of two numbers in Formula~\ref{eq:dot}. This step calculates the significance or 'weight' to be assigned to each segment of the signal, which we call attention scores $S_{atten}$. The weight matrices $W_{Q}$ and $W_{V}$ are shaped $(k \times k)$, and weight matrix $W_{K}$ is sized $(1 \times k)$. Given that the hidden states from LSTM has the dimension $(k \times T)$ and signal quality vector has the dimension $(1 \times T)$, the dimension of $S_{atten}$ is $(T \times T)$.
\begin{align}
S_{atten} = \frac{QK^T}{\sqrt{d_k}}
\label{eq:dot}
\end{align}
, where $d_{k}$ is used to normalize the attention scores. It is equal to $k$, the number of hidden units in the LSTM layer.
\begin{align}
W_{atten} = \text{SoftMax}(S_{atten})
\label{eq:weight}
\end{align}
In Formula~\ref{eq:weight}, a softmax function is then applied to convert these attention scores into a standardized form where they all add up to one, effectively scaling the attention across the signal. The standardized scores form an attention matrix $W_{atten}$, which, as outlined in Formula~\ref{eq:output}, is employed to calculate a context vector by weighting the value vectors according to these scores. 
\begin{align}
\text{Output} = W_{atten} \cdot V
\label{eq:output}
\end{align}
This output context vector is a refined summary of the entire PPG signal, adjusted according to the quality of the signal. It emphasizes the trustworthy segments of the signal while diminishing the influence of lower quality parts. Finally, the context vector goes through a fully connected layer and a sigmoid function, which together classify the PPG signal. The sigmoid function provides the probability of the signal belonging to a specific category, thus completing the process of integrating SQIs into PPG signal classification.

\section{Experiment}
\subsection{Dataset}
\label{sec:training_data}
In this study, we employed a comprehensive evaluation on multiple datasets. For training purposes, we used a large-scale dataset, and for evaluation, we utilized three additional external datasets, including one that is publicly available.  Additionally, we incorporated a smaller signal quality dataset with clean labels and detailed signal quality information, such as the presence and exact timings of low-quality segments. This dataset was used to train our signal quality assessment model. 

\subsubsection{Train dataset}
Our training dataset was sourced from 28,539 patients in a hospital environment, where continuous PPG signals were recorded from bedside monitors. These monitors flagged events such as atrial fibrillation (AF), premature ventricular contraction (PVC), and others. Our study focused on AF, PVC, and normal sinus rhythm (NSR), grouping PVC and NSR labels under Non-AF for a binary AF vs. Non-AF classification. The PPG signals were segmented into non-overlapping 30-second intervals, initially sampled at 240Hz (7,200 timesteps each), and then downsampled to 80Hz (2,400 time steps). The dataset was divided by patient IDs into training and validation sets. The training dataset included 13,432 patients with 2,757,888 AF and 3,014,334 Non-AF segments, while the validation set comprised 6,616 patients with 1,280,775 AF and 1,505,119 Non-AF segments. Given that the labels were automatically generated by monitors, some label noise is expected and estimated to be around 25\%. This estimate was derived by manually annotating a small sample of the dataset.

\subsubsection{Signal quality dataset}
\label{sec:signal_quality_data}
The signal quality dataset consists of 18,055 PPG segments from 13 stroke patients. The data collection settings for this dataset were consistent with those of our training dataset. A notable feature of this dataset is the detailed information on signal quality, including the presence and specific timings of segments with poor quality. The SQ model shown in Figure~\ref{fig:gen_frame} was trained using this dataset. 

\subsubsection{Test dataset}
\paragraph{Testset A (Public Source)}
Testset A is a public dataset from~\cite{torres2020multi}, featuring data from wrist-worn devices in ambulatory settings. Originally with 25-second segments, we augmented these to 30 seconds and resampled them to 2,400 timesteps. It contains 52,911 AF and 80,620 Non-AF samples from 163 patients, including those with AF and healthy individuals.
\paragraph{Testset B (Institution B)}
Sourced from institution B, Testset B was gathered using wrist-worn Samsung Simband devices from 98 ambulatory patients. We processed the data into 30-second segments with 2,400 timesteps each. The dataset includes 348 AF and 506 Non-AF segments, reviewed and annotated by medical professionals.
\paragraph{Testset C (Institution C)}
Testset C, collected from institution C, consists of fingertip PPG data from 126 hospital patients. We formatted the continuous signals into 30-second, non-overlapping segments, each downsampled to 2,400 timesteps. This dataset includes 38,910 AF and 220,740 Non-AF segments, annotated by cardiac electrophysiologists.

\subsection{Compared Model}
To evaluate the performance of our proposed model, we conduct a performance comparison against several baseline models, including ResNet-34 classifer~\cite{he2016deep}, LSTM model~\cite{yu2019review} and the hybird ResNet-34 and LSTM architecture. Moreover, we also include two recent AF detection models that are publicly accessible: the CMC model, which addresses the issues of inaccurate AF labels~\cite{10418255}, and anther model SiamAF, which is trained utilizing both PPG and ECG data but only utilizing PPG for inference~\cite{guo2023siamaf}. Given the presence of noisy label in our training dataset, as discussed in Section~\ref{sec:training_data}, we apply a variety of label noise mitigation techniques, including the strategy used in CMC study~\cite{10418255}, Symmetric Cross Entropy (SCE)~\cite{wang2019symmetric}, Joint Optimization Learning (JOL)~\cite{tanaka2018joint}, and Generalized Cross Entropy (GCE)~\cite{zhang2018generalized}. We utilized Area Under the Receiver Operating Characteristic curve (AUROC), F1 score, and Area Under the Precision-Recall Curve (AUCPR) as metrics to compare the performance of  SQUWA with several baselines.

\vspace{-0.4cm}
\subsection{Ablation Study}
To rigorously evaluate our SQUWA model, we performed a series of ablation studies, as outlined in Table~\ref{tab:ablation}. These experiments involved systematically removing certain components of the model to create different model variants to assess their individual contributions. We also removed different sized kernels from the signal compositor to see the effect of each. The NKS, NKM, and NKL variants were each modified by removing (N) the small (S), medium (M), and large (L) kernels (K), respectively. The NSC variant was designed to use the raw PPG signal directly, bypassing the composite signal (SC) as input. In the NFE variant, we took away the CNN that extracts features (FE) and let the composite signal be analyzed by the LSTM. For the NRN variant, we replaced the LSTM with a simpler global averaging layer with  SQI integration. The NSQ variant retained the CNN and LSTM combination but did not incorporate SQIs, to specifically assess the role of signal quality integration. Finally, to further explore the significance of SQI integration, we introduced the RSQ variant. This model has the complete SQUWA structure but is trained with randomly (R) generated SQIs, enabling us to evaluate the impact of accuracy of SQIs on model performance.

\begin{table*}[!ht]
\centering
\caption{Configuration for the ablation study, where 'x' denotes the inclusion of specific modules in the algorithm, and 'N/A' indicates the absence of such modules. SC means composite signal generation.}
\label{tab:ablation}
\resizebox{0.6\linewidth}{!}{%
\begin{tabular}{l|ccccccc}
\hline
\hline
               & \multicolumn{1}{l}{\textbf{sLx1 Conv}} & \multicolumn{1}{l}{\textbf{mLx1 Conv}} & \multicolumn{1}{l}{\textbf{xLx1 Conv}} & \multicolumn{1}{l}{\textbf{SC}} & \multicolumn{1}{l}{\textbf{CNN}} & \multicolumn{1}{l}{\textbf{LSTM}} & \multicolumn{1}{l}{\textbf{SQ-attention}}  \\ \hline
\textbf{SQUWA} & x                                      & x                                      & x                                                              & x                               & x                                & x                                & x                                          \\
\textbf{NKS}   & N/A                                    & x                                      & x                                                              & x                               & x                                & x                                & x                                          \\
\textbf{NKM}   & x                                      & N/A                                    & x                                                              & x                               & x                                & x                                & x                                          \\
\textbf{NKL}   & x                                      & x                                      & N/A                                                            & x                               & x                                & x                                & x                                          \\
\textbf{NSC}   & x                                      & x                                      & x                                                              & N/A                             & x                                & x                                & x                                          \\
\textbf{NFE}   & x                                      & x                                      & x                                                              & x                               & N/A                              & x                                & x                                          \\
\textbf{NRN}   & x                                      & x                                      & x                                                              & x                               & x                                & N/A                              & x                                          \\
\textbf{NSQ}   & x                                      & x                                      & x                                                              & x                               & x                                & x                                & N/A                                        \\ 
\textbf{RSQ}   & x                                      & x                                      & x                                                              & x                               & x                                & x                                & x                                          \\ \hline
\end{tabular}%
}
\end{table*}

\vspace{-0.4cm}
\section{Results}
The results in Table~\ref{tab:comp_results} show that SQUWA outperforms ResNet, LSTM, and the combined ResNet-34 + LSTM in terms of all three metrics across different test sets. CMC, a recent model tailored for AF detection using PPG signals known for its ability to manage label noise, is outperformed by our SQUWA model, despite SQUWA not being specifically designed to handle label noise. This is an important observation given that our training dataset is affected by label noise. Comparing SQUWA with SiamAF, a contemporary AF detection model trained using both PPG and ECG data, SQUWA shows competitive performance on Testsets B and C. However, it does not perform as well on Testset A. 
Furthermore, the last four rows of the table show SQUWA's performance when combined with strategies to address the label noise, such as the strategy used in the CMC model, as well as other methods like SCE, JOL, and GCE. Notably, these strategies improve SQUWA's performance, allowing it to surpass SiamAF on Testset A. For instance, SQUWA with JOL achieves a F1 score of 0.63 on Testset A, which is higher than the 0.61 F1 score from SiamAF. 

\begin{table*}[!ht]
\centering
\caption{Comparison of performance between the baseline models and SQUWA, evaluated using AUROC, F1 score, and AUCPR as metrics.}
\label{tab:comp_results}
\resizebox{0.95\textwidth}{!}{%
\begin{tabular}{ll|lll|lll|lll}
\hline
\hline
                                              &         & \multicolumn{3}{c|}{\textbf{Testset A}}                                                                & \multicolumn{3}{c|}{\textbf{Testset B}}                                                            & \multicolumn{3}{c}{\textbf{Testset C}}                                                                 \\ \hline
\multicolumn{1}{l|}{Model}                    & Data    & AUROC                          & F1                                & AUCPR                             & AUROC                          & F1                             & AUCPR                            & AUROC                            & F1                                & AUCPR                           \\ \hline
\multicolumn{1}{l|}{\textbf{ResNet-34}}       & PPG     & 0.54 ± 0.01                    & 0.4 ± 0.01                        & 0.25 ± 0.01                       & 0.63 ± 0.01                    & 0.6 ± 0.02                     & 0.53 ± 0.02                      & 0.68 ± 0.01                      & 0.28 ± 0.02                       & 0.23 ± 0.02                     \\
\multicolumn{1}{l|}{\textbf{LSTM}}            & PPG     & 0.48 ± 0.01                    & 0.23 ± 0.03                       & 0.21 ± 0.02                       & 0.41 ± 0.02                    & 0.47 ± 0.03                    & 0.39 ± 0.02                      & 0.57 ± 0.02                      & 0.18 ± 0.03                       & 0.12 ± 0.03                     \\
\multicolumn{1}{l|}{\textbf{ResNet-34 +LSTM}} & PPG     & 0.64 ± 0.01                    & 0.43 ± 0.01                       & 0.34 ± 0.02                       & 0.82 ± 0.01                    & 0.71 ± 0.01                    & 0.74 ± 0.01                      & 0.93 ± 0.01                      & 0.56 ± 0.01                       & 0.68 ± 0.01                     \\
\multicolumn{1}{l|}{\textbf{CMC}}             & PPG     & 0.76                           & 0.53                              & 0.6                               & 0.88                           & 0.75                           & 0.84                             & 0.94                             & 0.7                               & 0.73                            \\
\multicolumn{1}{l|}{\textbf{SiamAF}}           & PPG+ECG & {\ul \textbf{0.87}}            & 0.61                              & {\ul \textbf{0.72}}               & 0.9                            & 0.78                           & 0.86                             & 0.94                             & 0.73                              & 0.72                            \\
\multicolumn{1}{l|}{\textbf{SQUWA}}           & PPG     & \multicolumn{1}{r}{0.8 ± 0.01} & \multicolumn{1}{r}{0.56  ±  0.01} & \multicolumn{1}{r|}{0.63 ±  0.01} & \multicolumn{1}{r}{0.9 ± 0.01} & \multicolumn{1}{r}{0.8 ± 0.01} & \multicolumn{1}{r|}{0.85 ± 0.02} & \multicolumn{1}{r}{0.94 ±  0.01} & \multicolumn{1}{r}{0.73 ±  0. 01} & \multicolumn{1}{r}{0.75 ± 0.01} \\
\multicolumn{1}{l|}{\textbf{SQUWA + CMC}}     & PPG     & 0.82                           & 0.59                              & 0.66                              & 0.91                           & 0.8                            & 0.85                             & 0.95                             & 0.75                              & 0.79                            \\
\multicolumn{1}{l|}{\textbf{SQUWA + SCE}}     & PPG     & 0.84                           & 0.62                              & 0.7                               & 0.89                           & 0.79                           & 0.8                              & 0.94                             & 0.75                              & 0.79                            \\
\multicolumn{1}{l|}{\textbf{SQUWA + JOL}}     & PPG     & {\ul \textbf{0.87}}            & {\ul \textbf{0.63}}               & 0.7                               & {\ul \textbf{0.93}}            & {\ul \textbf{0.81}}            & {\ul \textbf{0.89}}              & 0.94                             & 0.74                              & 0.77                            \\
\multicolumn{1}{l|}{\textbf{SQUWA + GCE}}     & PPG     & {\ul \textbf{0.87}}            & 0.61                              & 0.71                              & 0.92                           & 0.8                            & 0.88                             & {\ul \textbf{0.95}}              & {\ul \textbf{0.76}}               & {\ul \textbf{0.8}}              \\ \hline
\end{tabular}%
}
\end{table*}

In the ablation study, we evaluate the performance of SQUWA against a series of its modified versions. As shown in Figure~\ref{fig:ablation}, the AUCPR metric is utilized to compare these models across three distinct datasets. From the results, it is apparent that the full SQUWA model generally outperforms its ablated counterparts. This suggests that all parts of the model work well together to make accurate predictions. For example, the version without the CNN for handling composite signals, named NFE, shows a notable decrease in performance across all datasets. This tells us that using a CNN to get features from PPG signals is important. When we look at how the model does with different kernel sizes in the signal compositor, we observe varying degrees of performance drop. The version with a medium-sized kernel, NKM, didn't do as poorly as the versions with small or large kernels. This means the medium-sized kernel might not be as important as the other two kernel sizes. However, the large kernel seems more important for the Testset B and Testset C, but not as much for Testset A, where it's about as important as the small kernel. Another interesting point is that the version NSQ without the signal quality integration, which is key for managing noisy data, had a significant drop in performance, especially in Testset A. The RSQ version, characterized by its randomly generated signal quality index, exhibited the poorest performance among all variants across the three datasets. These findings underscore the crucial role of signal quality. On the other hand, the NRN version didn't fall behind much from the full SQUWA model. Overall, these observations suggest that each component of the SQUWA model plays a vital role in its overall effectiveness, with certain components being particularly critical depending on the datasets evaluated.
\begin{figure*}[!ht]
\floatconts
  {fig:ablation}
  {\caption{Visualization of an ablation study showing the area under the precision-recall curve (AUCPR) across three test sets.}}
  {\includegraphics[width=0.75\linewidth]{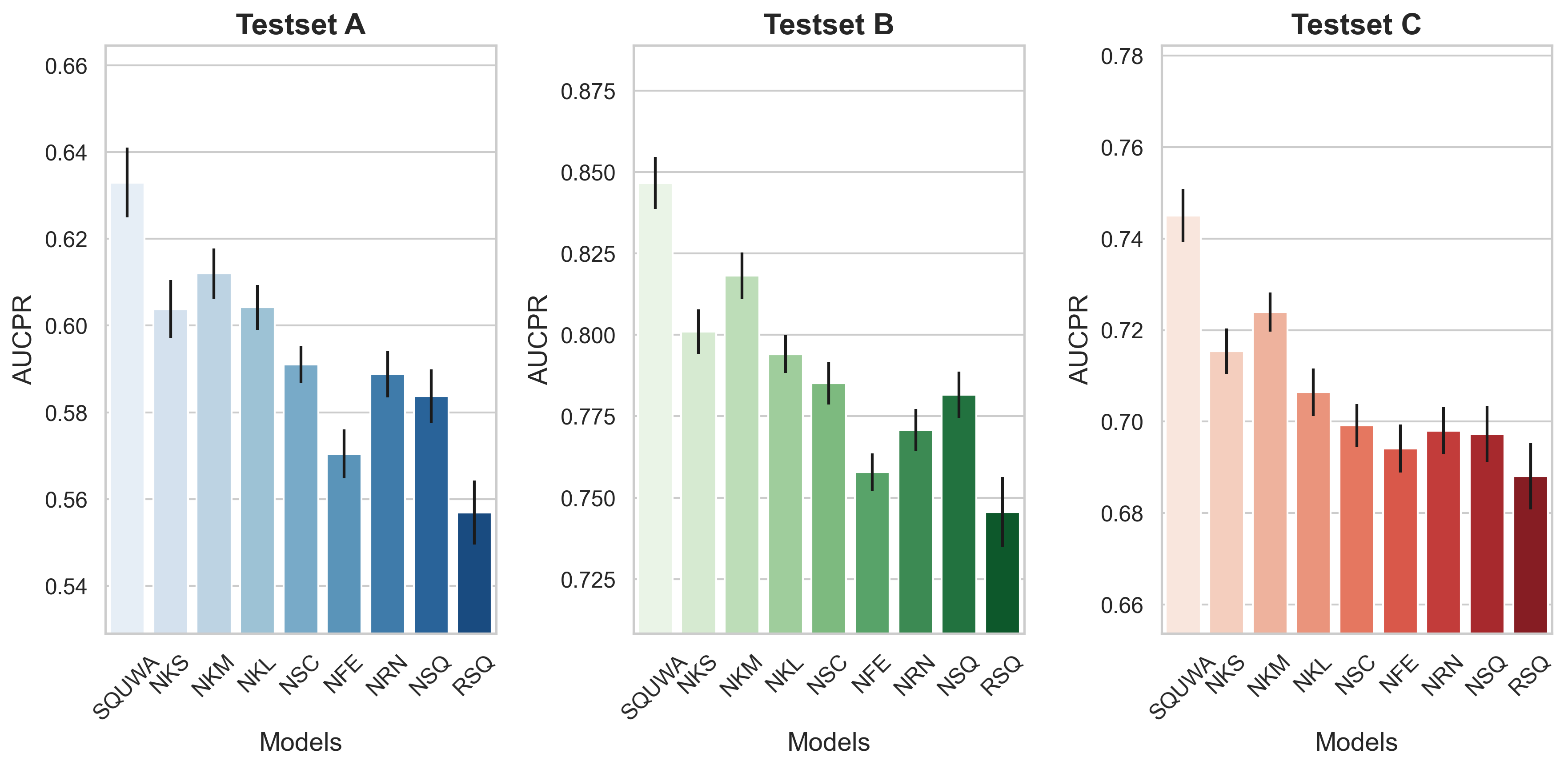}}
\end{figure*}

Figure~\ref{fig:combined_figure} presents the AUCPR scores in relation to the bad signal quality percentage for the SQUWA model and its variant NSQ that omits SQI integration. The y-axis measures the AUCPR, while the x-axis represents the percentage of the signal considered to be of bad quality based on SQIs from CAM. Each point on the graphs corresponds to AUCPR scores computed from signal samples with a bad quality percentage up to the indicated threshold. For the Testset A, as the threshold for bad signal quality increases, the AUCPR for SQUWA remains relatively stable and even shows a slight improvement, suggesting that the model is robust to varying signal quality. In contrast, the NSQ exhibits a decline in AUCPR, highlighting the model's dependency on higher signal quality for maintaining performance. In the case of the Testset B, the disparity between the two models becomes more pronounced. The SQUWA maintains its AUCPR scores substantially better as the signal quality worsens compared to the NSQ, which demonstrates a steeper drop. This indicates that the integration of SQI within SQUWA plays a significant role in preserving the model's performance under poor signal conditions. For the Testset C, both models show an increase in AUCPR as the threshold for bad signal quality increases, but SQUWA consistently outperforms NSQ. The SQUWA's AUCPR scores increase more steeply, reinforcing the benefit of SQI integration in managing lower-quality signals. Overall, these figures suggest that SQI integration is an important feature of the SQUWA model, helping to sustain performance across varying levels of signal quality, which is particularly evident when comparing to the NSQ variant that lacks this feature.

\begin{figure}[!ht]
\centering
\floatconts
  {fig:combined_figure}
  {\caption{Comparing AUCPR scores of the SQUWA model and its NSQ variant without signal quality integration. The comparison is shown on a graph with AUCPR on the y-axis and the percentage of signal with bad quality on the x-axis.}}
  {%
    \subfigure[Testset A]{\label{fig:sq_sub1}%
      \includegraphics[width=0.9\linewidth]{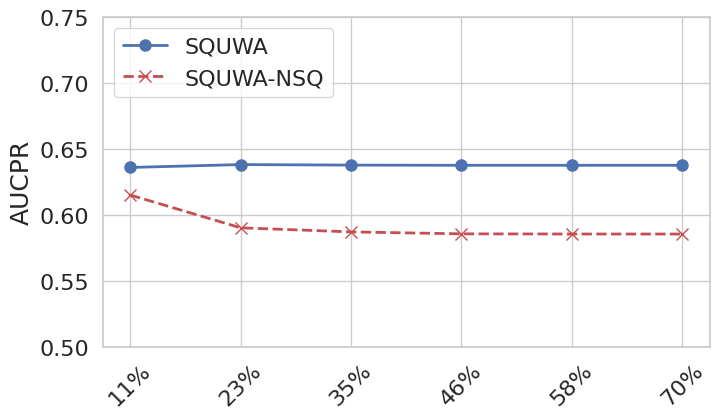}}%
    \qquad  
    \subfigure[Testset B]{\label{fig:sq_sub2}%
      \includegraphics[width=0.9\linewidth]{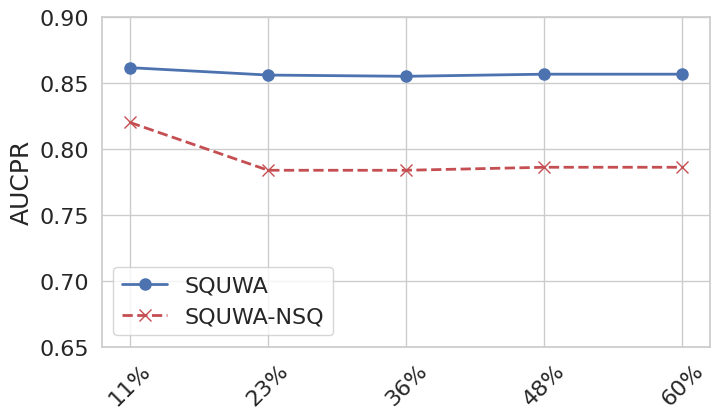}}%
    \qquad 
    \subfigure[Testset C]{\label{fig:sq_sub3}%
      \includegraphics[width=0.9\linewidth]{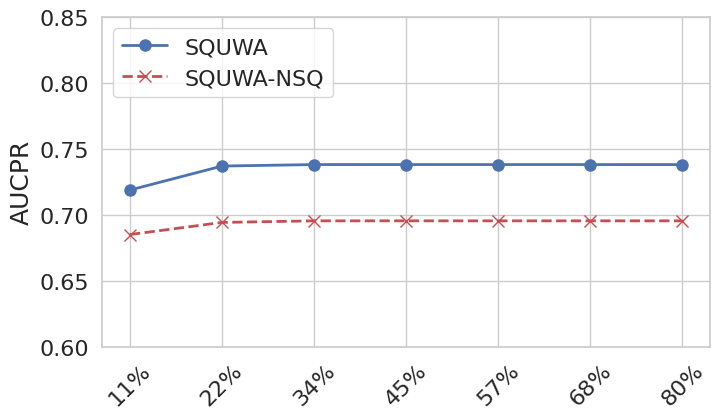}}
  }
\end{figure}

\begin{figure}[!h]
\floatconts
  {fig:atten_map}
  {\caption{This figure shows a Non-AF PPG signal with poor signal quality segments. The red area highlighted in the third figure indicates the portion of poor quality annotated by a person. The first and the second figures show the raw PPG signal and the signal after the composite signal generator in Figure~\ref{fig:signal_compositor}. The purple line of the third figure visualizes the signal quality, where a low value indicates poor signal quality}. Additionally, an attention heatmap shows the influence of each SQI component (x-aixs) on the LSTM's hidden states (y-axis), highlighting the impact of sequence segment quality on the LSTM representation.}
  {\includegraphics[width=0.95\linewidth]{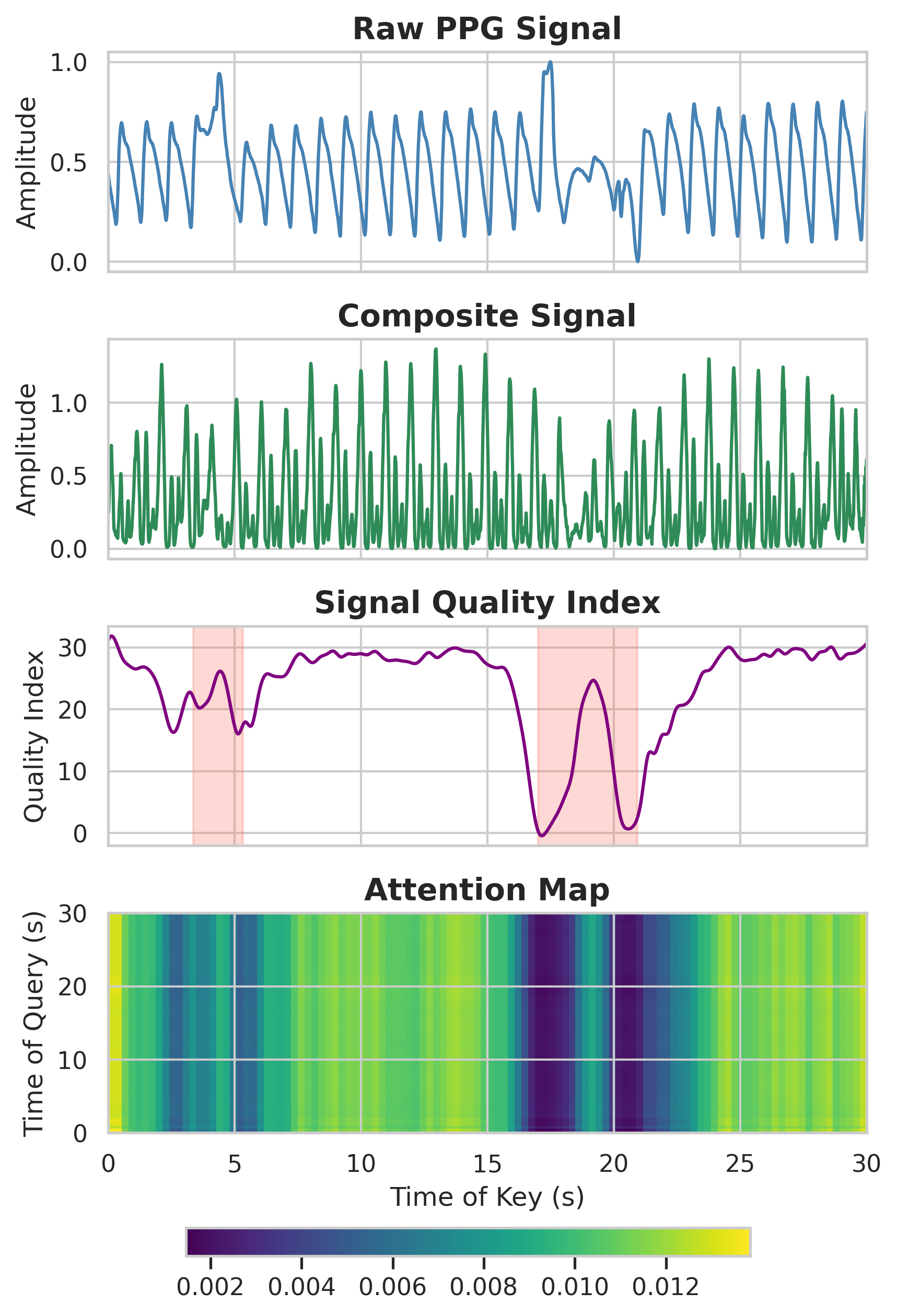}}
\end{figure}

As shown in Figures~\ref{fig:atten_map}, we applied the trained SQUWA model on the signal quality dataset mentioned in Section~\ref{sec:signal_quality_data}. Figure~\ref{fig:atten_map} illustrates an example of a Non-AF signal, with an additional AF example shown in Figure~\ref{fig:atten_map_af}. The first subplot depicts the amplitude variations corresponding to the raw PPG over time, while the second subplot visualizes the shape of the combined original signal and its derivatives after being passed through the composite signal generator. The composite signal shows more fine-grained details and contains a rich set of features than the raw PPG signal. In summary, the processed composite signal appears to extract a more comprehensive representation of the physiological characteristics as captured in the raw PPG signal and its temporal derivatives using varying-sized kernels. The purple curve represents the SQIs from SQ model, where a low value means worse signal quality. The red shaded areas represent periods manually annotated as having bad signal quality. It appears that the purple curve, which indicates the assessed signal quality, has valleys that align with the red shaded areas. This suggests that the SQIs are sensitive to the noise in the signal. The heatmap visualizes the attention score matrix and indicate how each time point in the SQIs (horizontal axis) influences each time point in the hidden states derived from the LSTM (vertical axis). Color intensity indicates the strength of the attention. A warmer color indicates higher attention weights, meaning that the SQI at that time point has a large influence on the hidden state from the LSTM at a given time. Each column on the horizontal axis corresponds to a moment in time for the SQI, and each row on the vertical axis corresponds to a moment in time for the hidden states derived from the LSTM. And all the elements of SQI and hidden states provide a sequential representation of the 30s input PPG signals. 
From the heatmap, we can observe that the bright yellow lines or spots indicate time points where the SQIs strongly influences the LSTM's hidden states. Conversely, the darker regions indicate time points that have less influence on the LSTM's hidden states, and these dark regions correspond with the red shaded areas in the SQIs, it suggests that the network is learning to disregard low-quality data.

\vspace{-0.4cm}
\section{Discussion \& Conclusion}
Detecting AF using PPG signals is crucial for use cases such as population-wide AF screening using wearable devices. However, a key challenge to realize such a potential of PPG is to account for impact of low signal quality on AF-detection sensitivity, which is important for screening AF at scale, and precision, which is critical to minimize untoward consequences of false detection.  In response to this challenge, we present the SQUWA neural network, which employs an attention mechanism designed to prioritize decision-making based on high-quality signal segments while mitigating the negative effects of corrupted segments, thereby enhancing the reliability of AF detection in PPG signals. Using SQUWA, there is no need to rely on using an arbitrary signal threshold to discard PPG siganls of poor quality. When assessing our method using three independent datasets not included in the training data collection, SQUWA outperformed traditional PPG models in classifying AF and Non-AF conditions. An ablation study revealed that the signal quality attention mechanism significantly boosts performance, with each component of the SQUWA model contributing positively to achieving performance seen in the full version of SQUWA. The attention maps validated our theory that the decision-making process favors segments with higher signal quality over those with lower quality. This attribute aligns with the insights of human domain experts, who can identify and tolerate noisy segments in the signals to a certain extent, while still making accurate judgments about AF by focusing on the segments of good quality. The issue of handling noisy time-series data is not unique to AF detection. Therefore, the principles underlying the SQUWA architecture could potentially be adapted for a range of other applications, such as human activity recognition and speech recognition, where similar data quality challenges exist.

It's important to note that alghtough the SQUWA model benefits from integrating the SQIs in the training process, it is not an entirely end-to-end system. The signal quality assessment component does require pre-training on a dataset with annotated signal quality, though the size of this dataset does not need to be substantial. Although we presented three test sets to demonstrate the robustness of SQUWA, more evaluations are needed to confirm its generalizability. Another limitation is that SQUWA is susceptible to inducing false negatives when artifacts obscure portions of signals where evidence AF is located. In practice, outputs from SQUWA processing consecutive 30-second strips can be further analyzed to enhance the model robustness for AF detection. 

\paragraph*{Institutional Review Board (IRB)}
The model development data was sourced from routine bedside monitor usage in the intensive care units at UCSF Medical Center (Institution A), under an IRB-approved waiver for written patient consents (IRB number: 14-13262). Testset B was collected from patients at Emory University Hospital undergoing AF ablation, who consented to wear a study device for PPG signal collection under IRB (00084629). Testset C was collected from routine bedside monitor usage in the acute care units at UCLA Medical Center with an IRB-approved waiver for written consents (IRB 10-000545).

\vspace{-0.4cm}
\acks{This research is supported by the National Institutes of Health (NIH) grant R01HL166233. It is incorporated into the DELTA project (Detecting and Predicting Atrial Fibrillation in Post-Stroke Patients), which has been registered on \href{https://classic.clinicaltrials.gov/ct2/show/NCT05795842}{clinicaltrials.gov}.}

\bibliography{chil-sample}

\appendix

\section{Implementation Details}
All experiments were conducted on a high-performance computing (HPC) cluster equipped with NVIDIA A100 and V100 GPUs. 
SQUWA was trained with these hyperparameters: batch size of 1024, and an Adam optimizer with a learning rate of 1e-4, using exponential decay. To prevent overfitting, we employed early stopping, stopping training if the validation loss did not improve for 10 epochs. For the composite signal generator, we used odd kernel sizes: 119 (about 1.5 seconds), 479 (about 6 seconds), and 799 (about 10 seconds), with a sampling frequency of 80Hz. Our architecture includes ResNet 34 for signal quality assessment and ResNet 18 for feature extraction, along with an LSTM with a hidden size of 64.

\begin{figure}[!h]
\floatconts
  {fig:atten_map_af}
  {\caption{This figure shows an AF PPG signal with segments of low quality.}}
  {\includegraphics[width=1.0\linewidth]{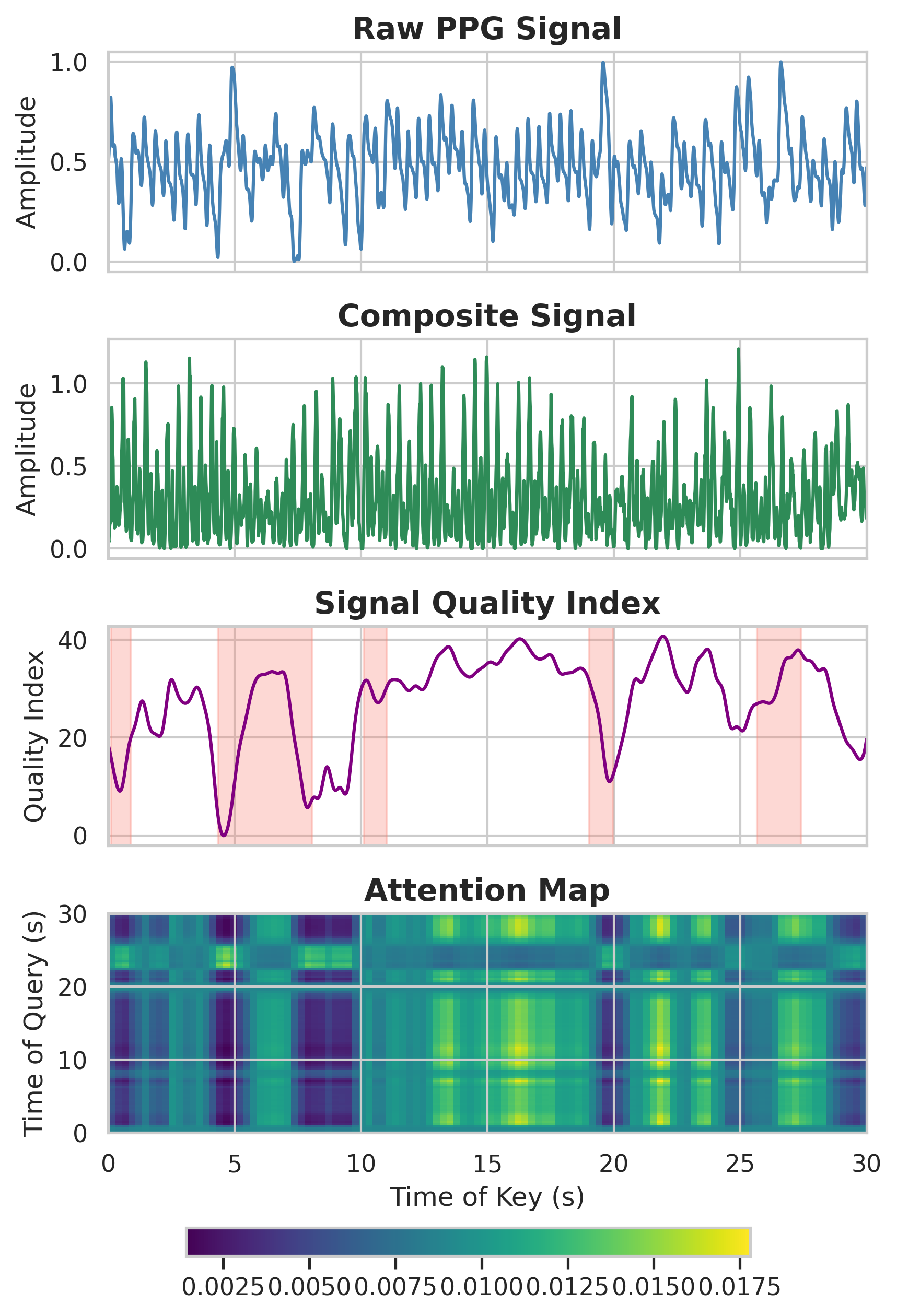}}
\end{figure}

\section{Attention Map of a AF Sample}
\label{apd:second}
Figure~\ref{fig:atten_map} shows a Non-AF sample, and we introduce an additional AF example in this section. The AF sample, as shown in Figure~\ref{fig:atten_map_af}, exhibits greater and more complex fluctuations compared to the Non-AF sample. But the generated signal quality index successfully identifies the corrupted segments, aligning with the human annotations marked by red shapes in the figure. The attention map reveals that the good quality sections receive higher weights compared to those that are corrupted. This demonstrates that our proposed method still works well for AF signals.

\section{Visualization of Test Samples}
\label{apd:third}

\begin{figure}[!ht]
\floatconts
  {fig:vis_exA}
  {\caption{Accurately classified AF and Non-AF samples from Testset A}}
  {%
    \subfigure[Non-AF]{\label{fig:testA_non}%
      \includegraphics[width=\linewidth]{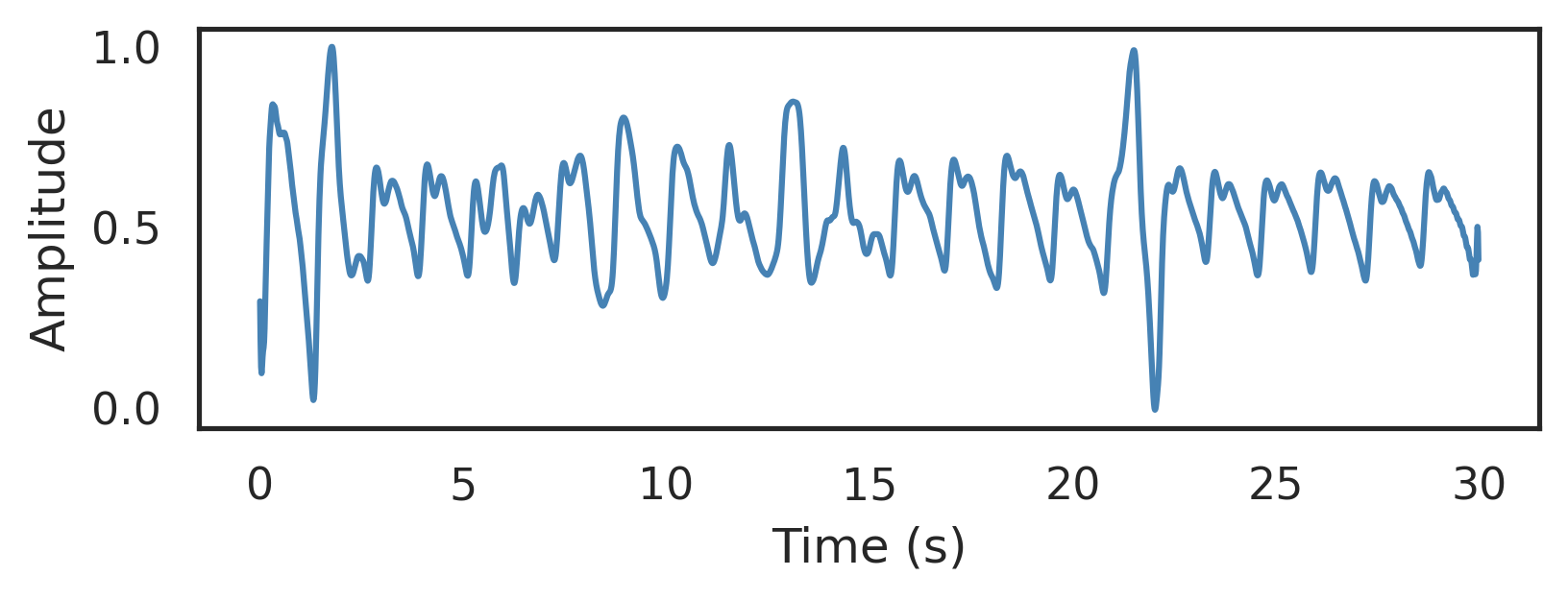}}
    \subfigure[AF]{\label{fig:testB_af}%
      \includegraphics[width=\linewidth]{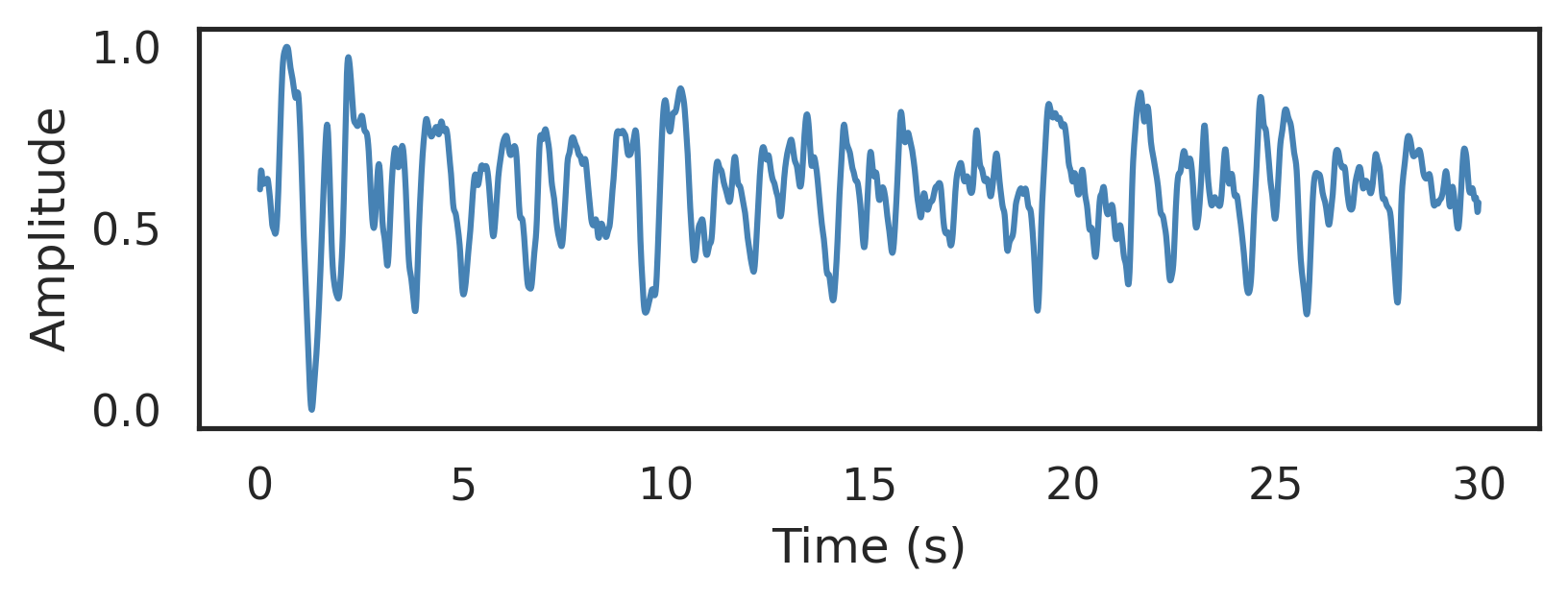}}
  }
\end{figure}

\begin{figure}[!ht]
\floatconts
  {fig:vis_exB}
  {\caption{Accurately classified AF and Non-AF samples from Testset B}}
  {%
    \subfigure[Non-AF]{\label{fig:testB_non}%
      \includegraphics[width=\linewidth]{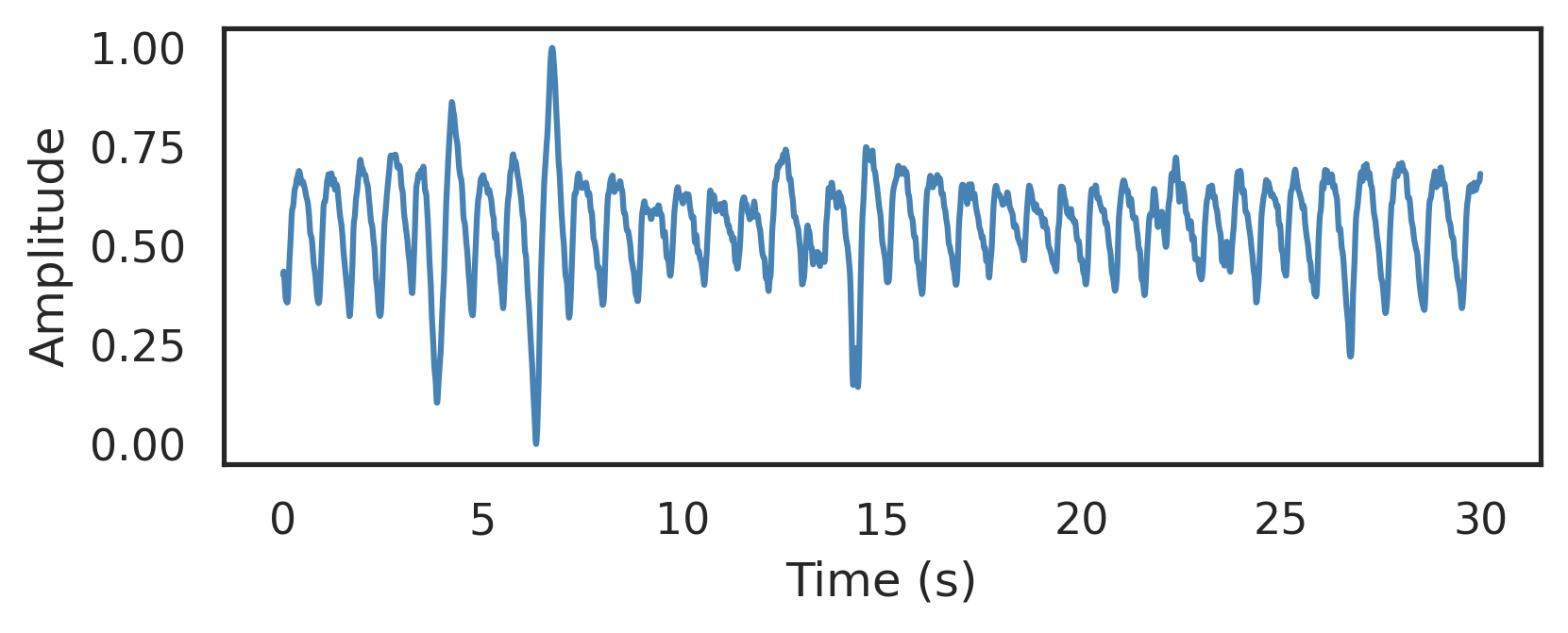}}
    \subfigure[AF]{\label{fig:testB_af}%
      \includegraphics[width=\linewidth]{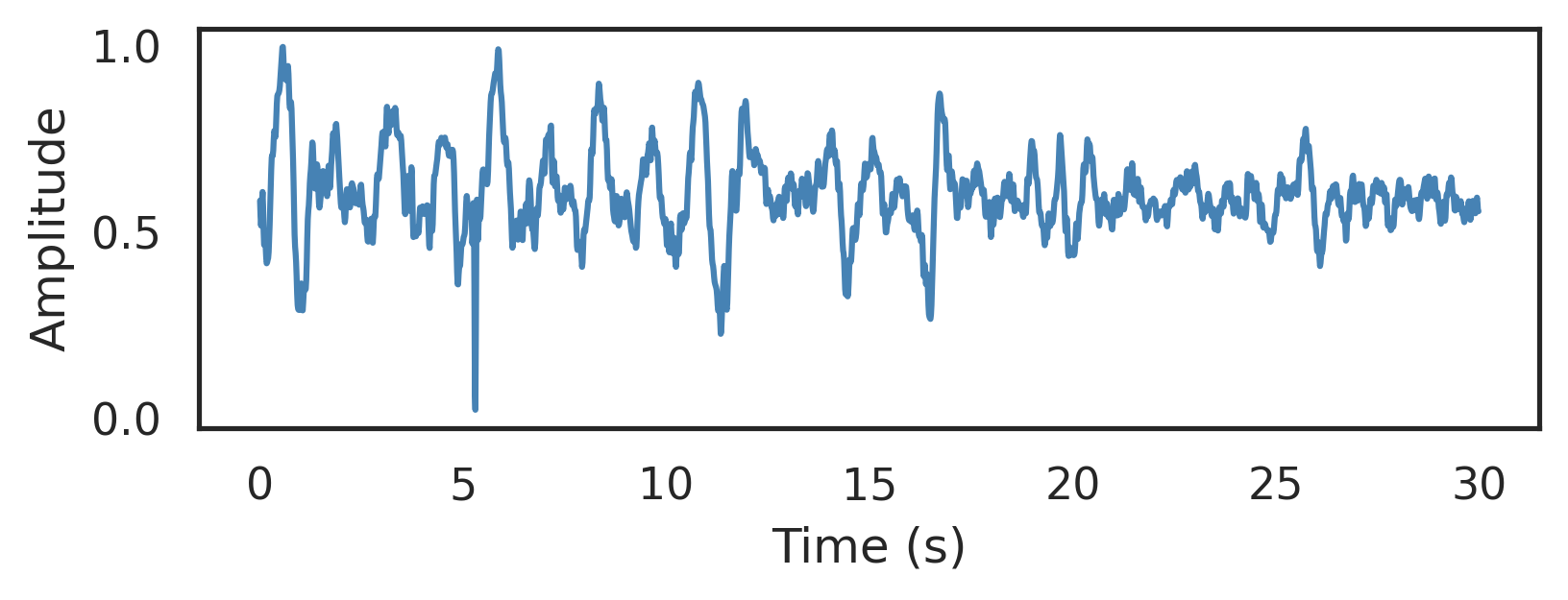}}
  }
\end{figure}

In this section, we provide additional instances of accurately classified AF and Non-AF samples for the three datasets: Testset A, as presented in Figure~\ref{fig:vis_exA}, Testset B, as presented in Figure~\ref{fig:vis_exB}, and Testset C, as presented in Figure~\ref{fig:vis_exC}. These EGM signals contain corrupted parts, demonstrating that SQUWA is able to accurately identify AF despite imperfections in the EGM signals.

\begin{figure}[t]
\floatconts
  {fig:vis_exC}
  {\caption{Accurately classified AF and Non-AF samples from Testset C}}
  {%
    \subfigure[Non-AF]{\label{fig:testB_non}%
      \includegraphics[width=\linewidth]{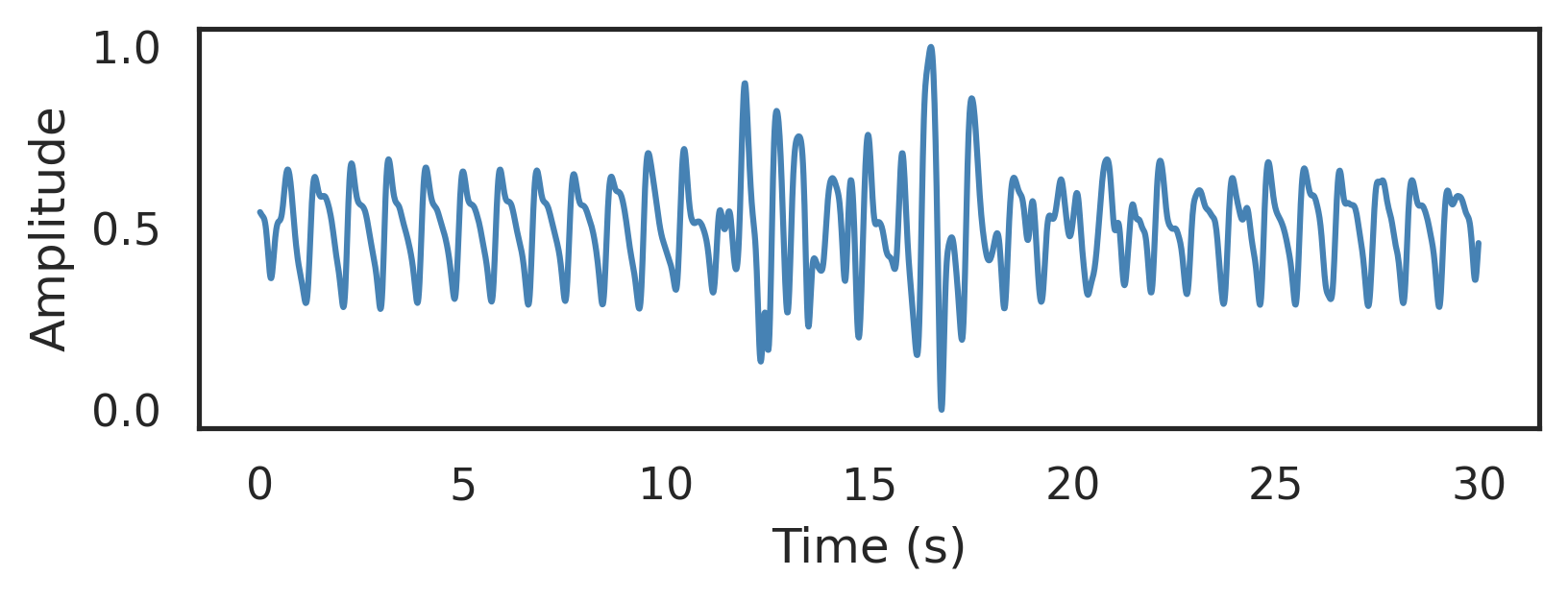}}%
    \qquad  
    \subfigure[AF]{\label{fig:testB_af}%
      \includegraphics[width=\linewidth]{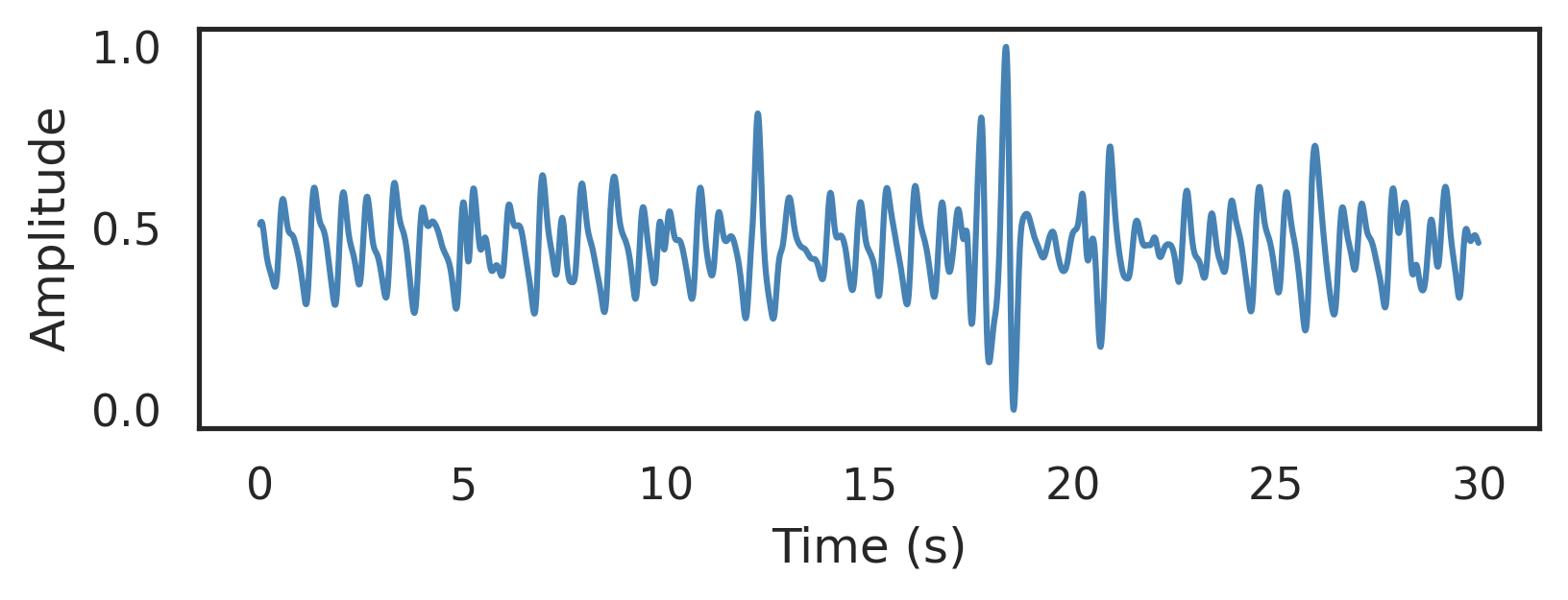}}%
  }
\end{figure}



\end{document}